\documentclass[sigconf]{acmart}
\AtBeginDocument{%
  }

\setcopyright{acmlicensed}
\copyrightyear{2025}
\acmYear{2025}
\acmDOI{XXXXXXX.XXXXXXX}
\acmConference[EASE '25]{International Conference on Evaluation and Assessment in Software Engineering}{June 17--20,
  2025}{Istanbul, Turkey}
\acmISBN{978-1-4503-XXXX-X/2018/06}

\usepackage{xcolor}
\usepackage{array}
\usepackage{listings}
\definecolor{lightgray}{gray}{0.9} 
\definecolor{codegreen}{rgb}{0,0.6,0}
\definecolor{codegray}{rgb}{0.5,0.5,0.5}
\definecolor{codepurple}{rgb}{0.58,0,0.82}
\definecolor{backcolour}{rgb}{0.95,0.95,0.92}

\begin{document}

\title{Test code generation at Ericsson using Program Analysis Augmented Fine Tuned LLMs}

\author{Sai Krishna, Balvinder Singh, Sujoy Roychowdhury, Giriprasad Sridhara}
\authornote{These authors contributed equally to this research}
\affiliation{%
  \institution{Ericsson}
    \country{India}
}




\author{Sourav Mazumdar}
\affiliation{%
  \institution{Ericsson}
    \country{India}
}

\author{Magnus Sandelin, Dimitris Rentas}
\affiliation{%
  \institution{Ericsson}
    \country{Sweden}
}


\author{Maciej Nalepa, Karol Sawicki, Jakub Gajda}
\affiliation{%
  \institution{Ericsson}
  \country{Poland}
}



\renewcommand{\shortauthors}{Bala et al.}


\begin{abstract}
We describe test code generation using Large Language Models (LLMs) in Ericsson. Our input is a test step in natural language (English) and our output is code (Java) which accomplishes the test step. We describe how straight forward prompting does not suffice and results in LLM assuming functions and signatures which are not present in the code repository. We then show how we alleviate the problem by a combination of Retrieval Augmented Generation (RAG) along with prompt engineering that expanded the simple prompt with additional contextual information using static program analysis. We then describe further improvements that we obtained by fine-tuning the underlying LLM. The fine tuning is done based on a custom designed prompt template which has pre-dependent classes, their public methods as well two exemplar outputs obtained from RAG. Our results establish that our fine tuned models help improve the correspondence or conformity with the original developer written test code as measured by the traditional metrics of F1-score based on the methods used in the generated code. Fine tuning of a 8x7b Mixture of Experts (MoE) model leads to an average improvement of 8\% over the base model and is comparable to the scores on a much larger 8x22b MoE model.
\end{abstract}





\maketitle

\section{Introduction}
\label{sec:intro}
Software quality assurance represents the bed rock of modern software. Quality assurance can be performed at multiple orthogonal levels such as static code analysis (using tools like Lint); manual or automated code review by expert developers; and software testing. Although each of these orthogonal techniques have their own importance in ensuring reliable software, testing arguably is the most important facet of software quality assurance. Software testing can be performed at multiple levels such as at an unit level (for example, a function or a method) or at the other end of the spectrum, the entire software (functional testing). Testing can also be done at an intermediate level such as component or multi component testing, where a component is a logical unit of a software.

Writing and maintaining tests has been shown to be a tedious and time consuming effort. Often, the same developers writing the actual software code have to write and maintain test code as well. Thus, to ameliorate the burden on developers, several automated approaches to generating test cases have been proposed.

Large Language Models have become the \emph{de facto} if not the \emph{de jure}
standard for automating several diverse tasks across the spectrum of software engineering. Unsurprisingly, testing has also seen the utilization of LLMs particularly for generating test code.

In this work, we describe our approach to automatically generating test cases for testing interactions between different components. We use Large Language models to automate the test code generation. Our input is a test step description in natural language such as ``connect to network element''. Our desired output is code in Java that accomplishes this natural language test step description. It typically consists of one or more Java statements, including conditionals such as an \emph{if-then-else} block and loops such as a \emph{for} loop.

We iterate through several approaches to generating test code. We start with simple prompts which lead to generated code with spurious objects and methods which meant that the generated code would not even compile. To alleviate this, we augment the prompt with static program analysis information and to further improve the generated code, we use Retrieval Augmented Generation. Finally, we use instruction fine tuning to achieve the desired level of correctness in the generated test code. 

The main contributions of this paper are as follows:

\begin{enumerate}
    \item A description of different approaches to test code generation using Large Language Models
    \item An evaluation of the different approaches
    \item A discussion of the relevant lessons that were learnt along the way
\end{enumerate}

The remainder of this paper is organized as follows: In Section~\ref{sec:approach} we describe our approach in greater detail followed by Section~\ref{sec:eval} where we delineate our evaluation. We portray the related work in Section~\ref{sec:rel} and present the future work and conclusion in Section~\ref{sec:conc}.

\begin{lstlisting}[caption={Sample Test Code Block to be generated. Some information has been modified / elided for confidentiality}, label=lst:sampleTestCode]
TestBegin("Reset parameter");
resetParameter(.., ..);
HelperClass.getInstance().update();
if (..) {
    doFunction();
} else {
    doOtherFunction();
}
TestEnd();
\end{lstlisting}

\section{Approach}
\label{sec:approach}

\subsection{Problem Definition and Dataset}\label{subsec:problem}

We consider proprietary Ericsson code repositories  for our study. Understanding these repositories involve a deep understanding of communication systems, signal processing and contain a large number of technical terminology and abbreviations.

Listing \ref{lst:sampleTestCode} shows a Test Code block which we need to generate via a LLM. The input to the system is expected to be the comment identified by the ``TestBegin'' phrase and the expected output is the code block within it.  The block ends with a \emph{TestEnd } statement. The ``TestBegin'' and ``TestEnd'' are two globally available methods in the repository indicating a component test case. We call the string within the ``TestBegin'' as Test Code Block Description
(TCBD).



In this study, we report the results on one Java repository consisting of {\color{black} hundreds of classes and thousands of test case  blocks}. {\color{black} Test code blocks have a mean of 10.5 and a Standard Deviation (SD) of 12.0 lines of code.} Test code blocks may have variables, object  instantiations of classes, calls to methods in the same or different class and logical blocks like \emph{if-else}, \emph{for} etc. 

Some of the challenges involved in the process are 
\begin{itemize}
    \item Code references to methods / variables in dependent libraries requiring the classes and methods to be available to the LLM 
    \item The need to use custom exceptions defined in certain utilities / common functions packages. These custom exceptions need to be made available to the LLM
    \item Functions with similar names different classes with different functionality
\end{itemize}


\begin{lstlisting}[caption={Sample IFT training prompt. Some information has been obfuscated / elided for confidentiality. Bold font has been added for easier reading.}, label=lst:sampleIFT, escapeinside={(*@}{@*)}]
    <(*@\textbf{s}@*)>[(*@\textbf{INST}@*)] You are an expert 5G network trace and test engineer and you are a Java programming expert. You are given a list of methods and example code blocks. Your task is to write a Java code block for a given test description
    <(*@\textbf{methods}@*)>
    Class Name: 	com.commonlibrary.x.y.z
    Method Names:	classX methodX()
        List<p> getSomething()
        Q getData()
        ..
    Class Name: 	com.p
    Method Names:	static String getContent(ClassA objectA)
        ..
    Class Name: 	com.helpers.p
    Method Names:	static String 
    ....
    </(*@\textbf{methods}@*)>
    <(*@\textbf{test\_description\_2}@*)>
    "Check that power is enabled"
    </(*@\textbf{test\_description\_2}@*)>
    <(*@\textbf{code\_block\_2}@*)>
            TestBegin("Check that power is enabled");
            String id = ClassB.getDetails(ClassC.getContent());
            assertTrue("Power is not " + ENABLED,
                    classA.methodB(objectP,id).isSuccessful());
            if ((m || n) && !j) {
                assertTrue("Power is not " + ENABLED + " here",
                        classC.methodD(objectQ, id).isSuccessful());
            }
            TestEnd();
    </(*@\textbf{code\_block\_2}@*)>
    <(*@\textbf{test\_description\_3}@*)>
    "Check that line is  enabled"
    </(*@\textbf{test\_description\_3}@*)>
    <(*@\textbf{code\_block\_3}@*)>
            TestBegin("Check that line is enabled");
            ...
            ...
            TestEnd();
    
    </(*@\textbf{code\_block\_3}@*)>
    Write code for the below test description.
    <(*@\textbf{test\_description\_1}@*)>
    "Check that device is enabled"
    </(*@\textbf{test\_description\_1}@*)>
     [/(*@\textbf{INST}@*)]
            TestBegin("Check that device is enabled");
            methodR();
            TestEnd();
    </(*@\textbf{s}@*)>
\end{lstlisting}

\begin{figure}[t!]
  \centering
\includegraphics[width=0.98\columnwidth]{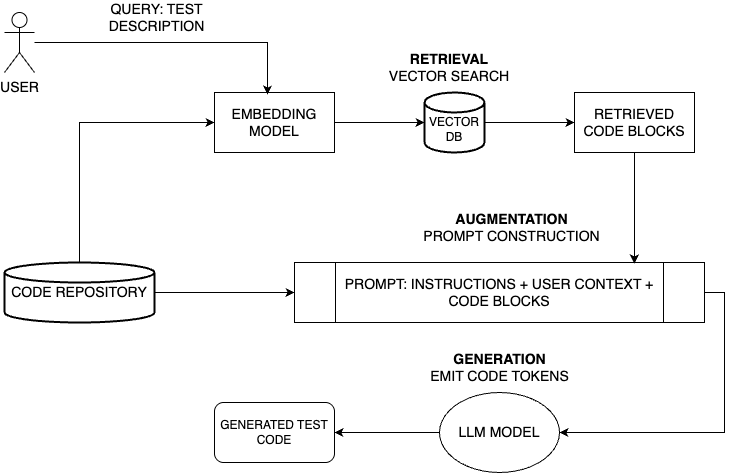}
  \caption{Schematic of our proposed RAG based fine tuned test code generation}
  \label{fig:schematic}
\end{figure}

We choose Mixtral 8x7b \cite{Mixtral8x7B} and Mixtral 8x22b \cite{Mixtral8x22B} as our LLMs - the choice of the models was dictated by a relatively large model with commercial friendly license in conjunction with legal experts.

\subsection{Experiment Details}\label{subsec:ExperimentDetails}

We describe here the different processing steps for our experiments.

\textbf{Simple Prompting} - A simple prompt asking a LLM to generate code by providing a TCBD leads it to assume function names and signatures which do not exist in the repository. Such code would obviously not compile and is not usable.

\textbf{Static Program Analysis} - We first do a static program analysis of the repository to identify class files, public methods with their fully qualified names and signatures and code comments / docstrings corresponding to the same. For each of the test code blocks, we identify the owning class, method and the TCBD.  Importantly, for each of our test code blocks we identify the fully qualified names of each of the methods invoked and their signature. 
We store the entire processed information in a graph with nodes being classes, methods and test code blocks and edges being relations of ownership and invocation between them.

\textbf{Embedding generation} We generate embeddings for both code block and TCBD by using a BERT based embedder finetuned for telecommunications domain and Ericsson internal data.

\textbf{Retrieval of similar code blocks} We retrieve similar code blocks by identifying TCBDs having a high cosine similarity with the query TCBD.

\textbf{Retrieval Augmented Generation} We retrieve two code blocks identified based on TCBD and ask the LLM to provide the test code.

\textbf{Prompt Construction} For our experimental setup, we create a custom prompt template. A sample (appropriately modified and elided for confidentiality) is given in Listing \ref{lst:sampleIFT}. Our prompt template consists of the following items
\begin{itemize}
    \item Instruction - This consists of the system prompt and the specific instruction for the LLM. This is identified by [INST]
    \item Based on the graph generated by the program analysis, for every TCBD we identify the java file it is part of and consider all the imports in the java file. This is identified by the tag ``<methods>'' in the prompt and has fully qualified class names and public methods and their signatures. For simplicity we consider protected methods in other classes as private too and do not include them in the prompt. All methods within the containing class of the TCBD are provided here. 
\item 
        Two TCBDs each followed by the corresponding test code block. These code blocks are obtained by retrieving similar code blocks similar to the way described for the RAG step. 
\item For the training and validation samples in addition to the instruction the ground truth  code block is provided        
\end{itemize}

Figure \ref{fig:schematic} shows a schematic of the end to end process for our custom prompt generation.


\begin{table}[]
    \resizebox{0.95\columnwidth}{!}{%
    \begin{tabular}
    {|p{0.5\columnwidth}|>{\raggedleft\arraybackslash}p{0.5\columnwidth}|}
    \hline
    \textbf{Parameter} & \textbf{Value} \\ 
    \hline
    Task Type & CAUSAL\_LM \\
    Rank (r) & 256 \\
    LoRA Alpha & 512 \\
    LoRA Dropout & 0.1 \\
    Bias & None \\
    \hline
    \end{tabular}%
    }
    \caption{PEFT parameters used for IFT}
    \label{tab:peft}
\end{table}

\textbf{Instruction Fine Tuning (IFT)} We use the prompt template defined above and the ground truth code block. The repository in question is dependent on multiple other repositories. For the methods section in the prompt, we identify classes and functions from all the repositories. There are over 10k classes across the repositories having tens of thousands of public methods. To train the model, we consider a context length of 10,000. We use Parameter Efficient Fine Tuning (PEFT) \cite{ding2023parameter} for the fine tuning process. The context length for our training pipeline is 10k tokens. Table \ref{tab:peft} has the parameters used in the PEFT training process. We only fine-tune the 8x7b model because of computational overheads both in training and in inference.


\section{Evaluation}
\label{sec:eval}

\subsection{Metrics}\label{subsec:metrics}

For evaluation, we consider the methods within the test code obtained by program analysis and compare with the methods in the generated code. Any method in the ground truth code not included in generated code is a false negative whereas any additional methods in generated code are false positives. Methods which appear in the generated and present in the ground truth are true positive. Based on this, we compute the F1 score of each of these code blocks.

While this metric does not capture if the logic is correct or not, it will capture if all the necessary method invocations are properly generated by the LLM. If this happens, then the developer would be able to modify the generated code in a much easier way.

\subsection{Results}\label{subsec:results}

Our results, using a simple prompt without providing any context, made the LLM assume function names and signatures and did not merit a quantitative evaluation. 

\begin{figure}[t!]
    \centering
\includegraphics[width=0.49\textwidth]{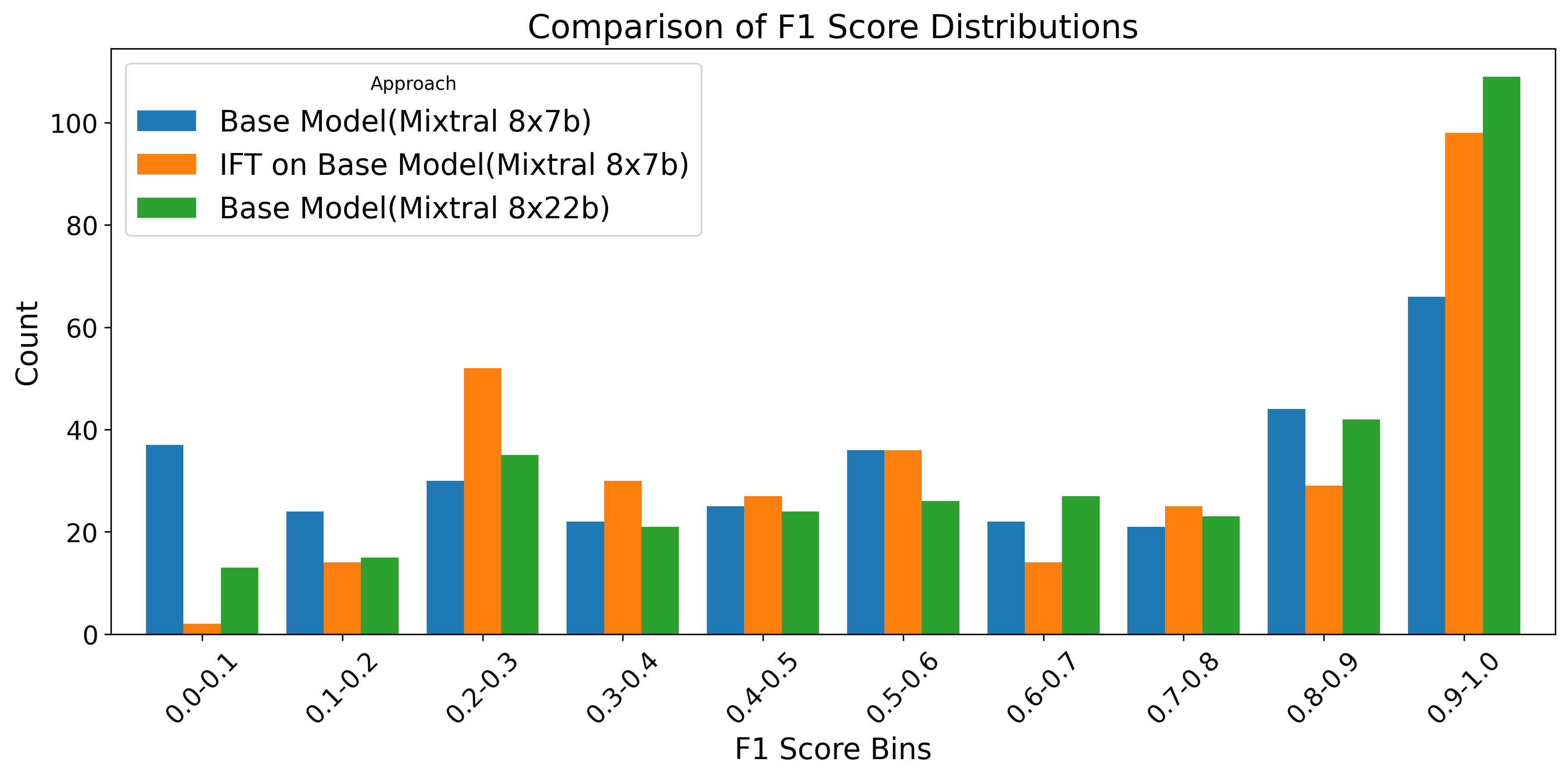}
    \caption{Comparative distribution of F1 scores for different models}
    \label{fig:results}
\end{figure}

\begin{table}[ht!]
    \centering
    \resizebox{0.8\columnwidth}{!}{%
    \begin{tabular}{|l|r|r|}
    \hline
    \textbf{Approach} & \textbf{F1 (Mean)} & \textbf{F1 (SD)} \\ 
    \hline
    Mixtral 8*7B & 0.55 & 0.33 \\
    \hline
    Mixtral 8x7B IFT & 0.63 & 0.30 \\
    \hline
    Mixtral 8x22B & 0.66 & 0.31 \\
    \hline
    \end{tabular}%
    }
    \caption{F1 score comparisons using different models}
    \label{tab:f1-scores}
\end{table}

The mean and standard deviations of the F1-scores from the different models is provided in Table \ref{tab:f1-scores}.  Figure \ref{fig:results} shows the histogram of F1-scores for the base (i.e. publicly downloaded) Mixtral 8x7b model, the base Mixtral 8x22b model and the IFT model as part of this study. We observe that IFT leads to a significant improvement on the results. In particular, extremely poor F1-scores ($<0.2$) are less than that of the larger public model.  We also note that the improvement in the highest F1-scores ($>0.9$) is very high and is comparable with that of the 3-times larger model.  A smaller model with similar performance available to a large development community leads to cost savings and  sustainability.

We observe from Table \ref{tab:f1-scores} and Figure \ref{fig:results} the variance of F1 scores is quite high. Poor scores were observed in cases where the ground truth had multiple conditional checks and custom exception calls - despite being provided in the context, LLM ignored many of these exception calls. Test blocks without any conditional loops scored very high, in general.


A simple approach such as prompting the LLM to generate Java code for a natural language description such as ``add network element to field'' does not work as the LLM hallucinates i.e. makes up methods and their signatures. For example, the LLM may generate code with a method call \emph{add} which while being superfically correct may be the incorrect method call, if the underlying operation is implemented using a synonymous name such as \emph{append}. This can be partially overcome by augmenting the prompt with program analysis information which provides the LLM with a list of potential or candidate methods that it can call to achieve the computational intent of the test step description. Other idiosyncrasies in the code that can only be remedied by providing the LLM with exemplar code already written by the developer for similar test steps. For example, in our code, there is a requirement to log the beginning and end of each test step. 
Thus, RAG was perforce needed for this task.
Instruction fine-tuning helps the LLM understand more about the underlying overall code base. 
Our experiments establish the benefits of a multi-staged approach involving fortifying the prompt with program analysis information, using Retrieval-Augmented Generation and finally performing instruction fine-tuning. 

\section{Related Work}
\label{sec:rel}

Earlier work has shown the potential of improving the quality and effectiveness of software development by the use of Large Language Models (LLM){\cite{sridhara2023chatgpt,hou2024large}}. In particular, the testing process can greatly benefit from  generated code as it reduces the time for an important, time-consuming component in the Software Development Life Cycle (SDLC). Unit test case generation has been looked at in various different works earlier \cite{zapkus2024unit,comadspaper}. Multicomponent testing (MCT), in between unit testing and functional testing, plays an important part in ensuring that different modules of a codebase work as expected. Some work \cite{lemner2024exploring,pan2024multi} has looked at component testing based on LLMs however this is a more evolving field \cite{li2025evaluating}.

\section{Conclusion and Future Work}
\label{sec:conc}
We described Java test code generation from a test step description in natural language (English) using Large Language Models. We showed that an approach that augmented the prompt with static program analysis and instruction fine-tuning yielded the best results in terms of the familiar F1 score. As dependencies can be extracted for any language, our methodology is applicable to other languages. Additionally, subject to the availability of more compute and data bigger models can be trained or higher PEFT ranks can be achieved for better performance. 

In future, we will work on the following: One limitation of our work is that we evaluate the generated code blocks only based on the F1 score of the methods invoked in the generated code block and the ground-truth code. Future work should look at comparing results by additionally considering the correctness of the logic implemented by the generated code.

Additionally, previous research has shown that continual pre-training \cite{yadav2023exploring} leads to better perfomance as the models can learn coding styles which may be unique to the company and a product / repository. Inclusion of continual pre-training and the improvement of the results is also a scope for future research.

Initial use of the solution in a focused user group has been positive, but evaluation of generative models like these would need a formal survey to study the benefits of the same to the developer community.  The productivity gains by such LLM based code generators need to be quantified and compared against the economic and environmental costs of training and hosting large models.

\bibliographystyle{ACM-Reference-Format}
\bibliography{main}


\begin{thebibliography}{11}


\ifx \showCODEN    \undefined \def \showCODEN     #1{\unskip}     \fi
\ifx \showISBNx    \undefined \def \showISBNx     #1{\unskip}     \fi
\ifx \showISBNxiii \undefined \def \showISBNxiii  #1{\unskip}     \fi
\ifx \showISSN     \undefined \def \showISSN      #1{\unskip}     \fi
\ifx \showLCCN     \undefined \def \showLCCN      #1{\unskip}     \fi
\ifx \shownote     \undefined \def \shownote      #1{#1}          \fi
\ifx \showarticletitle \undefined \def \showarticletitle #1{#1}   \fi
\ifx \showURL      \undefined \def \showURL       {\relax}        \fi
\providecommand\bibfield[2]{#2}
\providecommand\bibinfo[2]{#2}
\providecommand\natexlab[1]{#1}
\providecommand\showeprint[2][]{arXiv:#2}

\bibitem[AI(2023)]%
        {Mixtral8x7B}
\bibfield{author}{\bibinfo{person}{Mistral AI}.} \bibinfo{year}{2023}\natexlab{}.
\newblock \bibinfo{title}{Mixtral-8x7B-v0.1}.
\newblock \bibinfo{howpublished}{\url{https://huggingface.co/mistralai/Mixtral-8x7B-v0.1}}.
\newblock
\newblock
\shownote{Accessed: 2025-03-11}.


\bibitem[Community(2024)]%
        {Mixtral8x22B}
\bibfield{author}{\bibinfo{person}{Mistral Community}.} \bibinfo{year}{2024}\natexlab{}.
\newblock \bibinfo{title}{Mixtral-8x22B-v0.1}.
\newblock \bibinfo{howpublished}{\url{https://huggingface.co/mistral-community/Mixtral-8x22B-v0.1}}.
\newblock
\newblock
\shownote{Accessed: 2025-03-11}.


\bibitem[Ding et~al\mbox{.}(2023)]%
        {ding2023parameter}
\bibfield{author}{\bibinfo{person}{Ning Ding}, \bibinfo{person}{Yujia Qin}, \bibinfo{person}{Guang Yang}, \bibinfo{person}{Fuchao Wei}, \bibinfo{person}{Zonghan Yang}, \bibinfo{person}{Yusheng Su}, \bibinfo{person}{Shengding Hu}, \bibinfo{person}{Yulin Chen}, \bibinfo{person}{Chi-Min Chan}, \bibinfo{person}{Weize Chen}, {et~al\mbox{.}}} \bibinfo{year}{2023}\natexlab{}.
\newblock \showarticletitle{Parameter-efficient fine-tuning of large-scale pre-trained language models}.
\newblock \bibinfo{journal}{\emph{Nature Machine Intelligence}} \bibinfo{volume}{5}, \bibinfo{number}{3} (\bibinfo{year}{2023}), \bibinfo{pages}{220--235}.
\newblock


\bibitem[Hou et~al\mbox{.}(2024)]%
        {hou2024large}
\bibfield{author}{\bibinfo{person}{Xinyi Hou}, \bibinfo{person}{Yanjie Zhao}, \bibinfo{person}{Yue Liu}, \bibinfo{person}{Zhou Yang}, \bibinfo{person}{Kailong Wang}, \bibinfo{person}{Li Li}, \bibinfo{person}{Xiapu Luo}, \bibinfo{person}{David Lo}, \bibinfo{person}{John Grundy}, {and} \bibinfo{person}{Haoyu Wang}.} \bibinfo{year}{2024}\natexlab{}.
\newblock \showarticletitle{Large language models for software engineering: A systematic literature review}.
\newblock \bibinfo{journal}{\emph{ACM Transactions on Software Engineering and Methodology}} \bibinfo{volume}{33}, \bibinfo{number}{8} (\bibinfo{year}{2024}), \bibinfo{pages}{1--79}.
\newblock


\bibitem[Lemner et~al\mbox{.}(2024)]%
        {lemner2024exploring}
\bibfield{author}{\bibinfo{person}{Ludvig Lemner}, \bibinfo{person}{Linnea Wahlgren}, \bibinfo{person}{Gregory Gay}, \bibinfo{person}{Nasser Mohammadiha}, \bibinfo{person}{Jingxiong Liu}, {and} \bibinfo{person}{Joakim Wennerberg}.} \bibinfo{year}{2024}\natexlab{}.
\newblock \showarticletitle{Exploring the Integration of Large Language Models in Industrial Test Maintenance Processes}.
\newblock \bibinfo{journal}{\emph{arXiv preprint arXiv:2409.06416}} (\bibinfo{year}{2024}).
\newblock


\bibitem[Li et~al\mbox{.}(2025)]%
        {li2025evaluating}
\bibfield{author}{\bibinfo{person}{Yihao Li}, \bibinfo{person}{Pan Liu}, \bibinfo{person}{Haiyang Wang}, \bibinfo{person}{Jie Chu}, {and} \bibinfo{person}{W~Eric Wong}.} \bibinfo{year}{2025}\natexlab{}.
\newblock \showarticletitle{Evaluating large language models for software testing}.
\newblock \bibinfo{journal}{\emph{Computer Standards \& Interfaces}}  \bibinfo{volume}{93} (\bibinfo{year}{2025}), \bibinfo{pages}{103942}.
\newblock


\bibitem[Pan et~al\mbox{.}(2024)]%
        {pan2024multi}
\bibfield{author}{\bibinfo{person}{Rangeet Pan}, \bibinfo{person}{Myeongsoo Kim}, \bibinfo{person}{Rahul Krishna}, \bibinfo{person}{Raju Pavuluri}, {and} \bibinfo{person}{Saurabh Sinha}.} \bibinfo{year}{2024}\natexlab{}.
\newblock \showarticletitle{Multi-language unit test generation using llms}.
\newblock \bibinfo{journal}{\emph{arXiv preprint arXiv:2409.03093}} (\bibinfo{year}{2024}).
\newblock


\bibitem[Sridhara et~al\mbox{.}(2023)]%
        {sridhara2023chatgpt}
\bibfield{author}{\bibinfo{person}{Giriprasad Sridhara}, \bibinfo{person}{Sourav Mazumdar}, {et~al\mbox{.}}} \bibinfo{year}{2023}\natexlab{}.
\newblock \showarticletitle{Chatgpt: A study on its utility for ubiquitous software engineering tasks}.
\newblock \bibinfo{journal}{\emph{arXiv preprint arXiv:2305.16837}} (\bibinfo{year}{2023}).
\newblock


\bibitem[{Sujoy Roychowdhury} et~al\mbox{.}(2024)]%
        {comadspaper}
\bibfield{author}{\bibinfo{person}{{Sujoy Roychowdhury}}, \bibinfo{person}{{Giriprasad Sridhara}}, \bibinfo{person}{{A K Raghavan}}, \bibinfo{person}{{Joy Bose}}, \bibinfo{person}{{Sourav Mazumdar}}, \bibinfo{person}{{Hamender Singh}}, \bibinfo{person}{{Srinivasan Bajji Sugumaran}}, {and} \bibinfo{person}{{Ricardo Britto}}.} \bibinfo{year}{2024}\natexlab{}.
\newblock \showarticletitle{Static Program Analysis Guided LLM Based Unit Test Generation}.
\newblock \bibinfo{journal}{\emph{CODS COMADS}} (\bibinfo{year}{2024}).
\newblock


\bibitem[Yadav et~al\mbox{.}(2023)]%
        {yadav2023exploring}
\bibfield{author}{\bibinfo{person}{Prateek Yadav}, \bibinfo{person}{Qing Sun}, \bibinfo{person}{Hantian Ding}, \bibinfo{person}{Xiaopeng Li}, \bibinfo{person}{Dejiao Zhang}, \bibinfo{person}{Ming Tan}, \bibinfo{person}{Xiaofei Ma}, \bibinfo{person}{Parminder Bhatia}, \bibinfo{person}{Ramesh Nallapati}, \bibinfo{person}{Murali~Krishna Ramanathan}, {et~al\mbox{.}}} \bibinfo{year}{2023}\natexlab{}.
\newblock \showarticletitle{Exploring continual learning for code generation models}.
\newblock \bibinfo{journal}{\emph{arXiv preprint arXiv:2307.02435}} (\bibinfo{year}{2023}).
\newblock


\bibitem[Zapkus and Slotkien{\.e}(2024)]%
        {zapkus2024unit}
\bibfield{author}{\bibinfo{person}{Dovydas~Marius Zapkus} {and} \bibinfo{person}{Asta Slotkien{\.e}}.} \bibinfo{year}{2024}\natexlab{}.
\newblock \showarticletitle{Unit test generation using large language models: A systematic literature review}.
\newblock \bibinfo{journal}{\emph{Lietuvos magistrant{\k{u}} informatikos ir IT tyrimai: konferencijos darbai, 2024 m. gegu{\v{z}}{\.e}s 10 d.}} (\bibinfo{year}{2024}), \bibinfo{pages}{136--144}.
\newblock


\end{thebibliography}


\end{document}